\documentstyle[preprint,aps]{revtex}
\input epsf.tex
\def\DESepsf(#1 width #2){\epsfxsize=#2 \epsfbox{#1}}
\begin{document}
\preprint{\vbox{\hbox{}}}
\draft
\title{
$B\to \phi K$ and $B\to \phi X_s$ in the Heavy Quark Limit}
\author{Xiao-Gang He$^1$, J.P. Ma$^2$ and Chung-Yi Wu$^{1,3}$}
\address{
$^1$Department of Physics, National Taiwan University, 
Taipei\\
$^2$Institute of Theoretical Physics, Academia Sinica, Beijing\\
$^3$Department of Radiological Technology, Yuanpei Technical College,
Hsinchu
}
\date{ August, 2000}
\maketitle
\begin{abstract}
We study $B\to \phi K$ and $B\to \phi X_s$ decays in the heavy quark 
limit using perturbative QCD. The next leading order corrections 
introduce substantial modifications to the naive factorization 
results (more than 50\%). 
The branching ratio $Br(B\to \phi K)$ is predicted to be
in the range $(F^{B\to K}_1(m^2_\phi)/0.33)^2(3.5\sim 4.2)
\times 10^{-6}$ which 
is within the one $\sigma$ allowed region from the central value of
$6.2\times 10^{-6}$  measured by CLEO,
but outside the one $\sigma$ allowed region from the central value
of $17.2\times 10^{-6}$ measured by BELLE for reasonable $F_1^{B\to K}$. 
For the semi-inclusive decay $B\to \phi X_s$ we also  include initial
bound state effect in the heavy quark limit which decreases  
the branching ratio
by about 10\%. $Br(B\to \phi X_s)$ is predicted to
be in the range $(5.1\sim 6.3)\times 10^{-5}$.
\end{abstract}


\preprint{\vbox{\hbox{}}}


Recently CLEO and BELLE have measured the penguin
induced $\Delta S =1$ hadronic B decays with
$Br(B^-\to \phi K) = (6.4^{+2.5+0.5}_{-2.1-2.0}\times 10^{-6})$,
$Br(\bar B^0 \to \phi \bar K^0) = (5.9^{+4.0+1.1}_{-2.9-0.9}\times 10^{-6})$
from CLEO\cite{1}, and  
$Br(B^-\to \phi K^-) = (1.72^{+0.67+0.18}_{-0.54-0.18}\times 10^{-5})$ from
BELLE\cite{1a}.
Although the
central values do not agree with each other, they are consistent at
$2\sigma$ level. The branching ratios will
soon be determined with better precisions at CLEO, BABAR and BELLE.  
These decay modes are particularly interesting in the
Standard Model that they are purely due to penguins to the leading 
order\cite{2,3}
and therefore are sensitive to new physics at loop level\cite{4}. The neutral
decay mode also provides a model independent 
measurement for one of the KM unitarity
triangle parameter $\sin 2\beta$. The related semi-inclusive decay mode 
$B\to \phi X_s$
is also purely due to penguin\cite{2,5} and is 
sensitive to new physics at loop level. The branching ratio for this
decay although not measured at present, it 
will be measured in the near future at 
B factories. 
The above exclusive and semi-inclusive decays have been studied 
theoretically before with large errors\cite{2,3,5} that both the CLEO and 
BELLE measurements can be accommodated. 

Previous calculations for the branching ratios $Br(B\to \phi K)$ 
and $Br(B\to \phi X_s)$ are based on
naive factorization calculations. In these calculations, non-factorization
effects can not be calculated and are usually parameterized by an effective
color number and treated as a free parameter. There are also uncertainties
related to gluon virtuality in the penguin diagrams and dependence of
the renormalization scale. To have a better understanding of these decays,
it is necessary to carried out calculations in such a way that the problems
mentioned and others potential problems 
can be dealt with.
It has recently been shown that it is indeed possible in the heavy quark limit
to handle most of the problems mentioned in related B to two light mesons from 
QCD calculations\cite{6,7}. Several decays have been studied with 
interesting results\cite{8,9}. 
In this paper we will
follow the method developed in Ref.\cite{7} to carry out calculations for the
branching ratios for $B\to \phi K$ and $B\to \phi X_s$.

The effective Hamiltonian for charmless B decays with 
$\Delta S = 1$ is given by

\begin{eqnarray}
H_{eff} &=&{G_F\over \sqrt{2}}
\left \{ V_{ub} V^*_{us}(c_1 O_1 +c_2 O_2 + \sum_{i=3}^{11} c_i O_i)
+  V_{cb} V^*_{cs} \sum_{i=3}^{11} c_i O_i\right \}.
\end{eqnarray}
Here $O_i$ are quark and gluon operators and are given by

\begin{eqnarray}
&&O_1 = (\bar s_\alpha u_\beta)_{V-A} (\bar u_\beta b_\alpha)_{V-A},\;\;
O_2 = (\bar s_\alpha u_\alpha)_{V-A} (\bar u_\beta b_\beta)_{V-A},\nonumber\\
&&O_{3(5)} = (\bar s_\alpha b_\alpha)_{V-A}\sum_{q'} 
(\bar q^\prime_\beta q^\prime_\beta)_{V-(+)A},\;\;
O_{4(6)} = (\bar s_\alpha b_\beta)_{V-A}\sum_{q'} 
(\bar q^\prime_\beta q^\prime_\alpha)_{V-(+)A},\nonumber\\
&&O_{7(9)} = {3\over 2}(\bar s_\alpha b_\alpha)_{V-A}\sum_{q'} 
e_{q^\prime}(\bar q^\prime_\beta q^\prime_\beta)_{V+(-)A},\;\;
O_{8(10)} ={3\over 2} (\bar s_\alpha b_\beta)_{V-A}\sum_{q'} 
e_{q^\prime}(\bar q^\prime_\beta q^\prime_\alpha)_{V+(-)A},\nonumber\\
&&O_{11} = {g_s\over 8\pi^2} m_b \bar s_\alpha \sigma^{\mu\nu} G_{\mu\nu}^a 
{\lambda_a^{\alpha \beta}\over 2}(1+\gamma_5)b_\beta,
\end{eqnarray}
where $(V\pm A)(V\pm A) =\gamma^\mu(1\pm\gamma_5) \gamma_\mu(1\pm \gamma_5)$, 
$q^\prime = u,d,s,c,b$, $e_{q^\prime}$ is the electric charge number
of $q^\prime$ quark, $\lambda_a$ is the color SU(3) Gell-Mann matrix, 
$\alpha$ and $\beta$ are color indices, and
$G_{\mu\nu}$ is the gluon field strength.   

The coefficients $c_i$ are the Wilson Coefficients which have been 
calculated in
different schemes\cite{2,10}. In this paper we will use consistently the NDR 
scheme. The values of $c_i$ at $\mu \approx m_b$ GeV with the 
next-to-leading order (NLO) QCD corrections are given by\cite{10}

\begin{eqnarray}
&&c_1 = -0.185,\;\;c_2 = 1.082,\;\;c_3=0.014,\;\;c_4 = -0.035,\;\;
c_5=0.009,\;\;c_6 = -0.041,\nonumber\\
&&c_7= -0.002/\alpha_{em},\;\;c_8=0.054/\alpha_{em},\;\;
c_9=-1.292/\alpha_{em},\;\;c_{10}=-0.263/\alpha_{em},\;\;
c_{11} = -0.143.\nonumber
\end{eqnarray}
Here $\alpha_{em}=1/137 $ is the electromagnetic fine structure constant.

\noindent
{\bf The exclusive $B\to \phi K$ decay}

In the heavy quark limit, the decay amplitude due to a particular
operator can be represented in the
form\cite{7}

\begin{eqnarray}
<\phi K|O|B> = <\phi K|O|B>_{fact} [1+ \sum r_n \alpha_s^n +
O(\Lambda_{QCD}/m_b)],
\end{eqnarray}
where $<\phi K|O|B>_{fact}$ indicates the naive factorization result.
The parameter $\Lambda_{QCD}\approx 0.3$GeV is the strong interaction scale.
The second and third terms in the square bracket indicate
higher order $\alpha_s$ and $\Lambda_{QCD}/m_b$
corrections to the matrix elements. 
Including the next-leading order corrections and use information from 
Ref.\cite{9}, we have
the decay amplitude for $B\to \phi K$ in the heavy quark limit

\begin{eqnarray}
&&A(B\to \phi K)=
{G_F\over \sqrt{2}}C
<\phi|\bar s\gamma_\mu s |0> <K|\bar s \gamma^\mu b|B>,\nonumber\\
&&C=
V_{ub}V^*_{us}[a_3^u +a_4^u+a_5^u - {1\over 2}(a_7^u + a_9^u + a_{10}^u
)+ a_{10a}^u] \nonumber\\
&&\;\;\;\;\;\;
V_{cb}V^*_{cs}[a_3^c +a_4^c+a_5^c - {1\over 2}(a_7^c + a_9^c + a_{10}^c
)+ a_{10a}^c)].
\end{eqnarray}
We will use the notation
$<\phi| \bar s \gamma_\mu b | B > = m_\phi f_\phi \epsilon^\phi_\mu$
and $<K|\bar s \gamma^\mu b|B> = F_1^{B\to K}(q^2) (p_B^\mu + p_K^\mu)
+(F^{B\to K}_0(q^2)-F^{B\to K}_1(q^2)) (m_B^2-m_K^2)q^\mu/q^2$.

The coefficients $a_i^{u,c}$ are given by

\begin{eqnarray}
&&a_3^u=a^c_3 = c_3 + {c_4\over N} + {\alpha_s\over 4 \pi}
{C_F\over N} c_4 F_\phi,\nonumber\\
&&a_4^p = c_4 + {c_3\over N}
+{\alpha_s\over 4\pi} {C_F\over N}
\left [ c_3(F_\phi + G_\phi(s_s) + G_\phi(s_b))
+ c_1 G_\phi(s_p)\right. \nonumber\\
 &&\;\;\;\;+ \left .
(c_4+c_6) \sum_{f=u}^b G_\phi(s_f) + c_{11}G_{\phi,11}\right ],
\nonumber\\
&&a_5^u=a_5^c = c_5 +{c_6\over N} + {\alpha_s \over 4 \pi}
{C_F\over N} c_6(-F_\phi - 12),\nonumber\\
&&a_7^u = a^c_7 = c_7 + {c_8\over N} + {\alpha_s \over 4\pi}
{C_F\over N} c_8(-F_\phi -12),\nonumber\\
&&a_9^u = a_9^c = c_9 +{c_{10}\over N} + {\alpha_s\over 4\pi}
{C_F\over N} c_{10}F_\phi,\nonumber\\
&&a_{10}^u = a_{10}^c = c_{10} +{c_{9}\over N} + {\alpha_s\over 4\pi}
{C_F\over N} c_{9}F_\phi,\nonumber\\
&&a_{10a}^p=
{\alpha_s\over 4\pi} {C_F\over N}
\left [ (c_8+c_{10}) {3\over 2} \sum _{f=u}^be_f G_\phi(s_f)
+c_9 {3\over 2} (e_s G_\phi(s_s) + e_b G_\phi(s_b))\right ],
\end{eqnarray}
where $p$ takes the values $u$ and $c$, $N=3$ is the number of color,
$C_F = (N^2-1)/2N$, and $s_f = m^2_f/m_b^2$.
The other items are given by

\begin{eqnarray}
&&G_\phi(s) = {2\over 3} - {4\over 3} \ln{\mu\over m_b}
+ 4\int^1_0 dx \phi_\phi(x) \int^1_0 du u(1-u) \ln[s-u(1-u)(1-x)],\nonumber\\
&&G_{\phi,11} = -\int^1_0 dx {2\over 1-x} \phi_\phi(x),\nonumber\\
&&F_\phi = -12 \ln{\mu \over m_b} - 18 + f^I_\phi + f^{II}_\phi,
\nonumber\\
&&f^I_\phi = \int^1_0 dx g(x) \phi_\phi(x),\;\;
g(x) =  3{1-2x\over 1-x} \ln x -3i\pi,\nonumber\\
&&f^{II}_\phi = {4\pi^2\over N}
{f_K f_B\over F_1^{B\to K}(0) m_B^2}
\int^1_0 dz {\phi_B(z)\over z} \int^1_0 dx {\phi_K(x)\over x}
\int^1_0dy {\phi_\phi(y)\over y}.
\end{eqnarray}
Here $\phi_i(x)$ are meson wave functions. In this paper we will
take the following forms for them\cite{8},

\begin{eqnarray}
&&\phi_B(x) = N_B x^2(1-x)^2 Exp[-{m_B^2 x^2\over 2 \omega_B^2}],
\nonumber\\
&&\phi_{K,\phi}(x) = 6x(1-x),
\end{eqnarray}
where $N_B$ is a normalization factor satisfying $\int^1_0 dx \phi_B(x) = 1$.
Fitting various B decay data, $\omega_B$ is determined to be 0.4 GeV.

The above results are from genuine leading QCD calculation 
in the heavy quark limit. The
number of color should not be treated as an effective number, but has to be
3 from QCD. The results are renormalization scale independent. The problem
associated with the gluon virtuality $k^2 =(1-x) m^2_B$ 
in the naive factorization calculation
is also meaningfully treated by convoluting the $x$-dependence with the
meson wave functions in the functions $G(s,x)$. Also leading non-factorizable
is included (by the term proportional to $f_\phi^{II}$).
There are still uncertainties in the calculation, 
such as the form of the wave functions and 
the unknown $B\to K$ transition form factor $F_1^{B\to K}(q^2)$.  
However using wave functions obtained by fitting other data, the errors
can be reduced. In any case
calculations based on the
method used here is on more solid ground compared with previous calculations.

The decay rate can be easily obtained and is given by
\begin{eqnarray}
\Gamma(B\to \phi K) =
{G_F^2\over 32\pi} |C|^2 f_\phi^2 F_1^{B\to K}(m^2_\phi)^2 m^3_B
\lambda^{3/2}_{K\phi},
\end{eqnarray}
where $\lambda_{ij} = (1-m^2_i/m_B^2-m^2_j/m^2_B)^2 
- 4 m^2_i m_j^2/m_B^4$.

In our numerical calculations we will use the following values for the 
relevant parameters\cite{11}:
$m_b = 4.8$ GeV, $m_c = 1.4$ GeV, 
$V_{us} = 0.2196$, $V_{cb} = 0.0395$, $V_{ub}/V_{cb}
= 0.085$,
$f_\phi = 0.233$ GeV, 
$f_K = 0.158$ GeV, and
$f_B = (180 \pm 20)$ MeV. 
We keep the phase $\gamma$ to be a free parameter.
The results on the branching ratios are not sensitive to light quark masses.
We obtain the branching ratios for $B\to \phi K$ to be

\begin{eqnarray}
Br(B^-\to \phi K^-) &=&\left 
({F_1^{B\to K}(m^2_\phi)\over 0.33}\right )^2 (3.7\sim 4.2)\times 10^{-6},\nonumber\\
Br(\bar B^0 \to \phi \bar K^0) &=& \left ({F_1^{B\to K(m^2_\phi)}\over 0.33} \right )^2
(3.5\sim 4.1)\times 10^{-6}.
\end{eqnarray}

We have checked sensitivities on some of the
parameters.
The branching ratios are insensitive to the phase angle $\gamma$ because terms
proportional to $e^{-i\gamma}$ are suppressed by $|V_{ub}V_{us}^*/V_{cb}V_{cs}^*|$
which is about 1/50. The error associated with $V_{cb}$ is about 5\%. The
ranges given for the branching ratios have taken the dependence on $\gamma$
and the uncertainty in $V_{cb}$  
into account. 
We find that the NLO corrections to the matrix elements (terms proportional 
to $\alpha_s$ in $a_i$) to be significant. Without such NLO
corrections, 
the branching ratios are in the range of $(F^{B\to K}_1(m^2_\phi)/0.33)^2
(2.3\sim 2.5)\times 10^{-6}$.
The non-factorizable contributions (terms proportional to $f^{II}_\phi$) tend
to reduce the branching ratios at a few percent level.  
The from factor $F_1^{B\to K}$ is the least known parameter in the
calculations. 
There are several calculations for this parameter. Lattice calculation gives
$0.27\pm 0.11$\cite{12}, BSW model gives $0.38$\cite{13}, 
while light-cone calculation gives
$0.35\pm 0.05$\cite{14}. 
Using the average central value from these calculations, $F_1^{B\to K}(0)=0.33$,
 one finds that the
predicted branching ratios 
are closer to the averaged central value of the measurements from
CLEO than that from BELLE. 
To reach the CLEO central values, $F_1^{B\to K}$
needs to be around $0.42$ which is on the high value side from theoretical
calculations, 
while to reach BELLE central value a unreasonably large
value $0.72$ for $F_1^{B\to K}$ is needed. Precise measurements of these
modes may provide a good measurement of the form factor $F^{B\to K}_1$.
If a better understanding of the form factor $F^{B\to K}_1$ can be obtained
from other experimental measurements and from theoretical calculations in
the future, precise measurement of $B\to \phi K$ may provide us with 
important information about new physics beyond the SM.

\noindent
{\bf The semi-inclusive $B\to \phi X_s$ decay}

We will follow the procedures for semi-inclusive B decays described in Ref.\cite{15}
to study $B\to \phi X_s$.
The final state $X_s$ can be viewed as containing 
a perturbatively produced $s$ quark and some 
non-perturbatively produced state $X$
containing no strange number. Neglecting color octet contribution the
decay width for each of the helicity state $\lambda$ of $\phi(\lambda)$, 
at the leading order with light quark masses set to zero, 
can be written as 

\begin{eqnarray}
\Gamma_\lambda(B\to \phi X_s)
&=& {3\over 2} G_F^2 |\tilde C|^2 \int {d^3 k \over (2\pi)^3}
{d^4q\over (2\pi)^4 } 2\pi \delta(q^2) \int d^4x e^{iq\cdot x}
\nonumber\\
&\cdot&
<B|\bar b \gamma_\mu (1-\gamma_5)\gamma\cdot q\gamma_\nu (1-\gamma_5) b(x)
|B>\\
&\cdot& \sum_X <0|\bar s(0) \gamma^\mu (1-\gamma_5) s(0)|\phi(\lambda)+X>
<\phi(\lambda)+X|
\bar s(x) \gamma^\nu(1-\gamma_5) s(x)|0>,\nonumber
\end{eqnarray}
where the parameter $\tilde C$ is related to the Wilson Coefficients $c_i$.
In the vacuum saturation approximation,

\begin{eqnarray}
&&\sum_X <0|\bar s(0) \gamma^\mu (1-\gamma_5) s(0)|\phi(\lambda)+X>
<\phi(\lambda)+X|
\bar s(x) \gamma^\nu(1-\gamma_5) s(x)|0>\nonumber\\
&&\approx 
<0|\bar s(0) \gamma^\mu (1-\gamma_5) s(0)|\phi(\lambda)>
<\phi(\lambda)|
\bar s(x) \gamma^\nu(1-\gamma_5) s(x)|0>.
\end{eqnarray}

In this approximation the color octet contributions are automatically
neglected. We will work with this approximation to estimate the 
branching ratio for $B\to \phi X_s$. This approximation is consistent 
with the assumption made in the previous section if color octet is 
neglected. If one cuts the $\phi$ momentum
to be above 2 GeV or so, the contributions are dominated by the
effective two body decay $b\to \phi s$. In this case $\tilde C$
is similar to $C$
but with $f^{II}_\phi$ set to be zero.
In principle terms proportional to $f^{II}_\phi$ also contribute.
However this contribution is small and can be neglected.
This is because that in the semi-inclusive decay
only $\phi$ in the final state is specified. When the constraint of having
$K$ in the final state is relaxed, the term corresponding to $f_\phi^{II}$ 
leads to a three body decay. Requiring the identified hadron in the final state
to be hard limits the phase space\cite{15} and 
results in a small contribution from 
$f_\phi^{II}$ compared with other contributions.

If the b quark mass is infinitively large, $Br(B\to \phi X_s)$ is
equal to $Br(b\to \phi s)$. However due to initial b quark bound state
effect there are corrections\cite{16}. This correction is included
in the factor $<B|\bar b(0) \gamma_\mu (1-\gamma_5) \gamma\cdot q
\gamma^\nu (1-\gamma_5) b(x)|B>$. Following the discussions in
Ref.\cite{16} we obtain $1/m^2_b$ correction factor,

\begin{eqnarray}
\Gamma(B\to \phi X_s)
\approx {G_F^2 f^2_\phi m^3_b\over 16\pi} |\tilde C|^2
(1+{7\over 6}{\mu^2_g\over m_b^2}
-{53\over 6} {\mu^2_\pi\over m_b^2}),
\label{semi}
\end{eqnarray}
where

\begin{eqnarray}
\mu^2_g &=& <B|\bar h {1\over 2} g_s G_{\mu\nu} \sigma^{\mu\nu} h|B>,
\nonumber\\
\mu^2_\pi &=& - <B|\bar h D^2_T h|B>.
\end{eqnarray}
Here the field $h$ is related to $b$ by $b(x)
= e^{-im_b v\cdot x}$ $ \{
1+ i\gamma\cdot D_T/2m_b + v\cdot D \gamma\cdot D_T/4m^2_b
-(\gamma \cdot D_T)^2/8m^2_b\} h(x) + O(1/m_b^3) + 
(\mbox{terms for anti-quark})$. 
$D^\mu_T = D^\mu - v^\mu v\cdot D$ with $v$ being the four velocity of the
$B$ meson satisfying $v^2 =1$ and $D^\mu = \partial^\mu + i g_s G^\mu(x)$.

Analysis of spectroscopy of heavy hadrons
and QCD sum rule calculations 
give\cite{17}, $\mu^2_g \approx 0.36$ GeV$^2$ and $\mu^2_\pi \approx (0.3 \sim 
0.54)$ GeV$^2$. We will use $\mu^2_g = \mu^2_\pi = 0.36$ GeV$^2$ for
numerical calculations. The initial state $1/m_b^2$ correction reduces the
branching ratio by about 10\%. The branching ratio for
$B\to \phi X_s$ is predicted to be

\begin{eqnarray}
Br(B\to \phi X_s) = (5.1\sim 6.3)\times 10^{-5}.
\end{eqnarray}
This prediction, as in the case for exclusive decays, is insensitive
the phase angle $\gamma$. The NLO corrections enhance the branching ratio
significantly, similar to the exclusive decay cases.

The expression for the semi-inclusive decay in eq.(\ref{semi}), on the face
of it, has less parameters (no dependence on $F_1^{B\to K}$) 
compared with the exclusive branching ratios
discussed earlier.  One might think that the prediction for 
$Br(B\to \phi X_s)$ is more certain compared with the exclusive cases.
However, one should be careful about this because in the calculation we have
only included color singlet and the $^3S_1$ $s \bar s$ bound state 
contribution. There may be other contributions such as color octet and
other $S$ and $P$ wave 
states from $s \bar s$. These contributions are in general
smaller than the contributions already considered. One can not rule
out significant enhancement at the present.  However, we can view the 
color singlet result as leading contribution which gives a good order of
magnitude estimate of the semi-inclusive decay $B\to \phi X_s$.

In conclusion, 
we have studied $B\to \phi K$ and $B\to \phi X_s$ decays in the heavy quark 
limit using perturbative QCD. 
We found
that
the next leading order corrections 
introduce substantial modifications to the leading native factorization 
results (more than 50\%). 
The branching ratio $Br(B\to \phi K)$ is predicted to be
in the range $(F^{B\to K}_1(m^2_\phi)/0.33)^2(3.5\sim 4.2)
\times 10^{-6}$ which 
is within the one $\sigma$ allowed region from the central value of
$6.2\times 10^{-6}$  measured by CLEO,
but outside the one $\sigma$ allowed region from the central value
of $17.2\times 10^{-6}$ measured by BELLE for reasonable $F_1^{B\to K}$. 
For the semi-inclusive decay $B\to \phi X_s$ we also  included initial
bound state effect in the heavy quark limit which decreases  
the branching ratio
by about 10\%. $Br(B\to \phi X_s)$ is predicted to
be in the range $(5.1\sim 6.3)\times 10^{-5}$.
Future experimental data will provide us with more information about these 
decays and about method based on QCD improved factorization approximation.

This work was supported in part by NSC
under grant number NSC 89-2112-M-002-016.

\end{document}